\documentclass[%
 reprint,
 amsmath,amssymb,
 aps, prl,
]{revtex4-1}

\usepackage{graphicx}
\usepackage{dcolumn}
\usepackage{bm}
\usepackage{xcolor}

\begin{document}

\preprint{APS/123-QED}

\title{Criticality at finite strain rate in fluidized soft glassy materials}

\author{Magali Le Goff}
\email{magali.le-goff@univ-grenoble-alpes.fr}
\author{Eric Bertin}
\author{Kirsten Martens}%

\affiliation{%
Universite Grenoble Alpes, Laboratoire Interdisciplinaire de Physique, CNRS, F-38000 Grenoble, France
}


\begin{abstract}
We study the emergence of critical dynamics in the steady shear rheology of fluidized soft glassy materials.
Within a mesoscale elasto-plastic model accounting for a shear band instability, we show how an additional noise can induce a transition from phase separated to homogeneous flow, accompanied by critical-like fluctuations of the macroscopic shear rate. 
Both macroscopic quantities and fluctuations exhibit power law behaviors in the vicinity of this transition, consistent with previous experimental findings on vibrated granular media. 
Altogether, our results suggest a generic scenario for the emergence of criticality when shear weakening mechanisms compete with a fluidizing noise.
\end{abstract}

\maketitle

Dense disordered matter, e.g.\ in form of emulsions, foams, colloidal and granular materials, are known to display both solid-like and fluid-like features in response to applied deformation or stresses.
As a result they exhibit unusual stress-strain curves and complex rheological behaviours, leading to several interesting out-of-equilibrium transitions, which are accompanied by intermittent dynamics \cite{bonn2017yield,nicolas2018deformation}.
In the static yielding, materials undergo a transition upon increase of an externally applied deformation. They evolve from an elastic regime at small deformations, to a plastic flow-regime after reaching the so-called static yield stress \cite{varnik2004study}. The nature of this transition in the transient regime, potentially leading to strongly intermittent dynamics \cite{combe2000strain, karmakar2010statistical}, has been recently investigated in the context of non-equilibrium phase transitions \cite{jaiswal2016mechanical,leishangthem2017yielding, ozawa2018random,popovic2018elastoplastic}.
A second well studied transition in this context is the dynamic yielding transition, which is concerned with the steady flow regime at a vanishing but finite imposed shear rate $\dot\gamma$. In this quasi-static driving regime, the materials exhibit a finite dynamic yield stress, that can differ from the above defined static one \cite{varnik2004study}. The emerging critical dynamics in the steady flow of soft glassy materials in the vicinity of the dynamic yield transition has been extensively studied \cite{bailey2007avalanche, talamali2011avalanches, lin2014scaling,liu2016driving,aguirre2018critical}.

In this work we shall consider yet another, much less investigated non-equilibrium phenomenon, that emerges for systems exhibiting a non-monotonic flow curve, corresponding to a discontinuous dynamic yielding transition \cite{becu2006yielding,coussot2010physical,martens2012spontaneous}.
In this case the material separates in a flowing and a non-flowing region for strain rates smaller than a threshold value $\dot\gamma_c$. At a qualitative level, such a transition is reminiscent of equilibrium discontinuous phase transitions, such as the liquid-gas transition. It is well-known at equilibrium that by tuning temperature, a line of discontinuous transition may end with a critical point, like the liquid-vapor critical point. Back to the flow transition, this suggests that by tuning a control parameter
the discontinuous transition may generically end with a critical point, possibly located at a finite value of the shear rate \cite{porte1997inhomogeneous}.
This scenario has been confirmed in a recent experimental work on sheared and vibrated granular media by Wortel and co-workers \cite{wortel2016criticality}.
In this experiment, mechanical vibrations fluidize the granular packing at low shear stress and, upon a critical vibration magnitude, induce a transition from a non-monotonic to a monotonic flow curve, 
accompanied by critical-like fluctuations of the macroscopic strain rate.

Beyond the specific case of granular matter, we expect a finite shear-rate critical point to appear as soon as a soft glassy system exhibits both a non-monotonic flow curve and a fluidization mechanism.
In granular systems, non-monotonic flow curves have been traced back to frictional sliding contacts \cite{wortel2014rheology,degiuli2017friction},
and fluidization results from external mechanical vibration \cite{d2003observing,Caballero-Robledo2009,jia2011elastic,hanotin2012vibration,wortel2014rheology,lieou2015stick,pons2015mechanical,wortel2016criticality,degiuli2017friction}.
In soft frictionless systems, non-monotonic flow curves may result for instance from a local softening due to long restructuring times after a plastic rearrangement of particles \cite{coussot2010physical, martens2012spontaneous}. Fluidization may also result (apart from mechanical vibration) from local processes such as coarsening in foams  \cite{cohen2004origin}, or of active origin \cite{mandal2016active,tjhung2017discontinuous,matoz2017nonlinear}.

In this Letter, we explore this generic scenario for the emergence of a critical point in the framework of elasto-plastic models for the flow of soft, frictionless glassy materials \cite{picard2005slow,nicolas2018deformation}, which can be tuned to exhibit a non-monotonic flow curve \cite{picard2005slow,martens2012spontaneous}.
By adding an external source of noise in the model to generate a fluidization mechanism, we obtain in this generic minimal model a finite shear-rate critical point ending a line of discontinuous flow transition.
We characterize in details the scaling properties of the shear rate and of its fluctuations close to the critical point. We find in particular that while some of the critical exponents take simple mean-field values, the exponents characterizing the divergence of the correlation length and time take non-standard values that cannot be easily understood from an equilibrium analogy, even taking into account the presence of long-range interactions.

\begin{figure*}[th]
\centering
\includegraphics[width=1.9\columnwidth, clip]{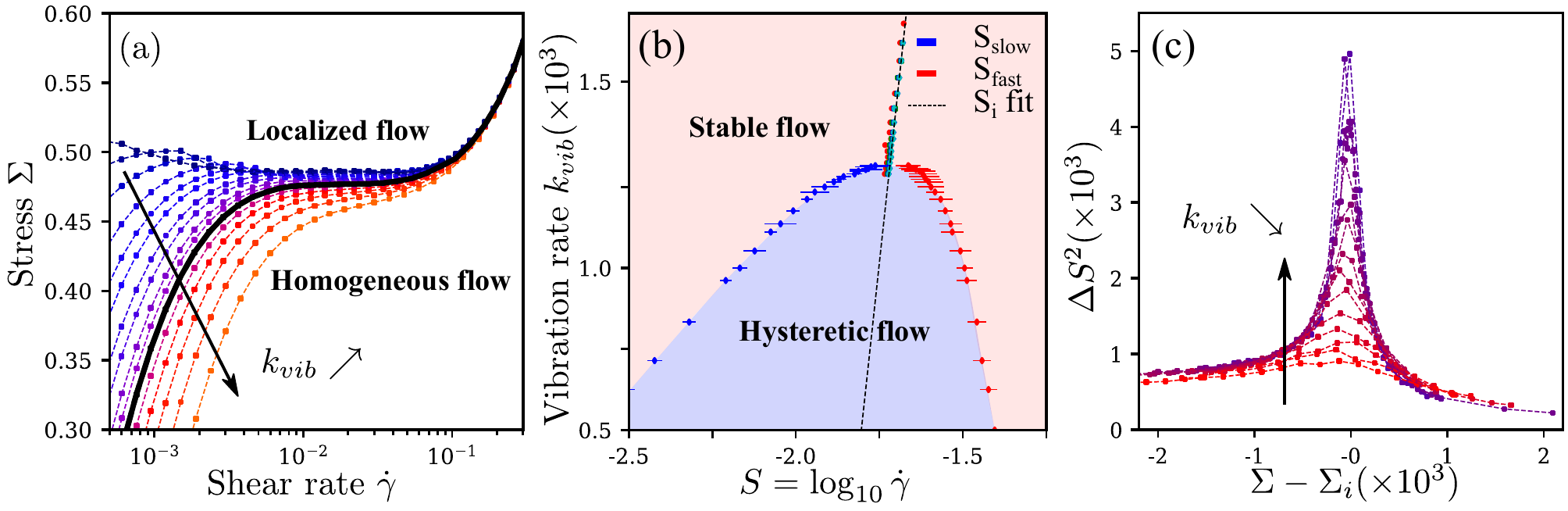}
\caption{\textbf{(a) Flow curves for various vibration rates $k_{vib}$} using a shear-rate-controlled driving protocol ($L = 256$). The thick black line indicates the transition from non monotonic to monotonic flow curves.
\textbf{(b) Regimes of stable and hysteretic flow} determined using a stress-controlled driving protocol ($L = 1024$). Blue and red data points indicate the limits of stability of the stable slow and fast flowing branches and the blue shadowed region depicts the regime of hysteretic flow (two coexisting solutions). \textbf{(c) Fluctuations of the logarithm of the macroscopic shear rate} measured using a stress-controlled protocol in the regime where $k_{vib}>k_{vib}^c$, as a function of the distance to the stress at the inflexion point of each flow curve ($L = 256$).}
\label{fig1_Phenomenology}
\end{figure*}

\paragraph*{Elasto-plastic model:}

Coarse-grained elasto-plastic models (EPM) provide a generic framework for the rheology of soft glassy materials (see \cite{nicolas2018deformation} and references therein). 
In EPM, the stress increases uniformly across the system under a uniform driving, either by controlling the strain rate or the stress in the system.
When the local stress overcomes a threshold value, $\sigma_y$, particles rearrange locally \cite{argon1979plastic} in a plastic fashion, causing a local relaxation of the stress and an elastic response of the surrounding solid-like material. The elastic propagation kernel is described using the Eshelby propagator \cite{eshelby1957determination}, with an asymptotic power-law decay $\sim 1/r^d$ ($d$ being the spatial dimension) and a quadrupolar symmetry. 

\begin{figure*}[t]
\centering
\includegraphics[width=1.9\columnwidth, clip]{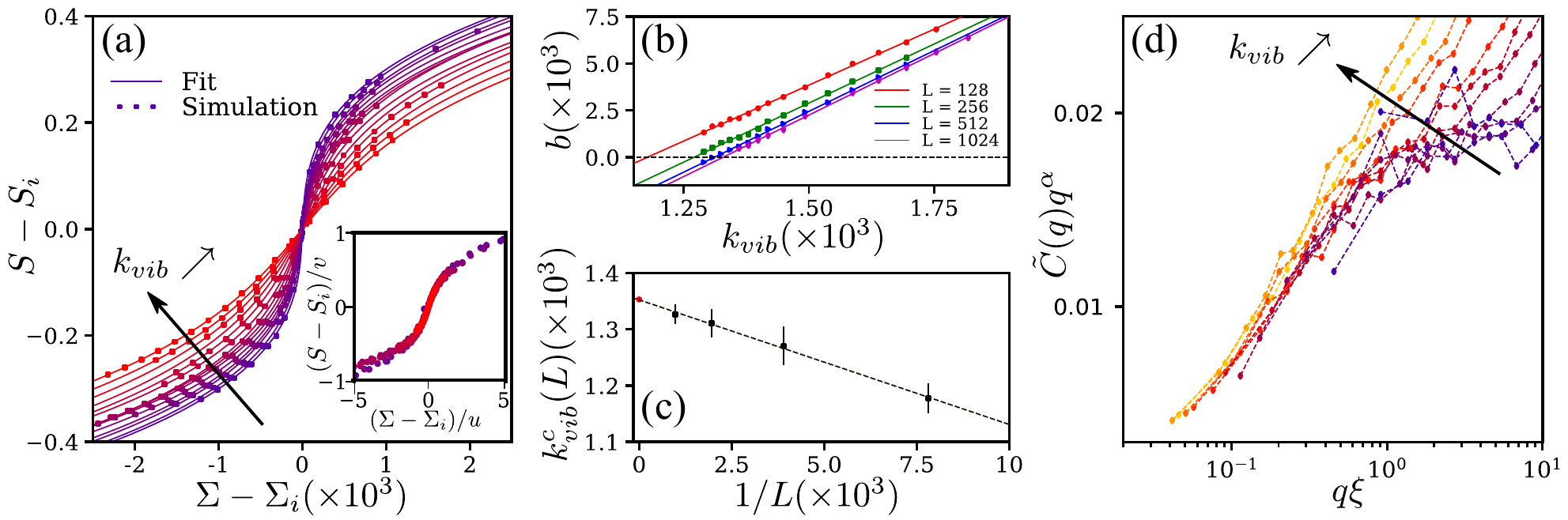}
\caption{\textbf{(a) Landau type expansion fit for $k_{vib}>k_{vib}^c$.}
$S-S_i$ as a function of $\Sigma-\Sigma_i$, $S_i$ and $\Sigma_i$ being the coordinates of the inflexion point of each flow curve. Dots are simulation points for $L=256$ (averaged over a strain $\gamma>2.10^4$) and solid lines are fits to equation \ref{fit_Landau}. Inset: data collapse using $u=(b/3a)^{1/2}$ and $v=(4b^3/0.027a)^{1/2}$, 
\textbf{(b) Susceptibility.} Inverse susceptibility $b$ as a function of $k_{vib}$ for $L=128,256,512,1024$ and linear fits to locate the value $k_{vib}^c(L)$ for which $b(k_{vib}^c,L)=0$.
\textbf{(c) Critical point location.}  $k_{vib}^c(L)$ as a function of the inverse system size $L^{-1}$ and linear extrapolation (dashed line) to locate the critical point $k_{vib}^c(\infty)$ in the infinite system limit (red dot).
 \textbf{(d) Spatial correlations.} Fourier transform of the autocorrelation of the local shear rate field (for  $L=1024$)  normalized by a power law dependence $\tilde C(q)q^\alpha$ as a function of $q \xi$ with  $\xi = \bigl((k_{vib}-k_{vib}^c)/{k_{vib}^c}\bigr)^{\nu}$ with $\nu =1 $ enabling data collapse for small values of $q$.}
\label{fig2_flow_curve_scaling}
\end{figure*}
Our modeling approach is built upon the recently introduced stress-controlled driving EPM \cite{liu2018creep}. 
We coarse-grain an amorphous medium onto a square lattice of size $L \times L$ where the mesh size is adjusted to the typical cluster size of rearranging particles undergoing a plastic rearrangement (the lattice indices $i$,$j$ represent the discretized coordinates along $x$ and $y$ directions respectively). We assume for these local plastic transformations the same geometry as the globally applied simple shear, i.e., we consider a scalar model.
Conceptually we decompose the total deformation of each node into a local plastic strain $\gamma_{ij}^{pl}$
and an elastic strain $\gamma_{ij}^{el}$. 
Further we decompose also the local stress into two parts, $\sigma_{ij}=\sigma^{ext}+\sigma_{ij}^{int}$, where $\sigma^{ext}$ is the externally applied uniform stress, and $\sigma_{ij}^{int}$ encodes the stress fluctuations resulting from the interactions between plastic regions, as described by:
\begin{equation}
\sigma_{ij}^{int}=\mu\underset{i'j'}{\sum}G_{ij,i'j'}^{*}\gamma_{i'j'}^{pl} 
\end{equation}

The interaction kernel $G$, of Eshelby's type, reads in Fourier space:
$\tilde{G^{*}}({\mathbf q}) = -4 \frac {q_x^2 q_y^2}{q^4}$ for $\mathbf{q} \neq 0$
and $\tilde{G^{*}}({\mathbf 0}) = 0$ 
so that $\sigma_{ij}^{int}$ describes the local stress fluctuations in a macroscopically stress-free state. 
Applying a macroscopic stress $\sigma^{ext}$ induces a uniform shift of the local stress without altering internal fluctuations. 
The local dynamics is expressed as:
\begin{equation}
\frac{d}{dt}\gamma_{ij}^{pl}=n_{ij}\frac{\sigma_{ij}}{\mu\tau}=n_{ij}\frac{\sigma^{ext}+\sigma_{ij}^{int}}{\mu\tau}
\end{equation}
with $\mu$ the elastic modulus, $\tau$ a mechanical relaxation time setting the time units of the model and $\frac{d}{dt}\gamma_{ij}^{pl}=\frac{n_{ij}\sigma_{ij}}{\mu\tau}$ the strain rate produced by a plastic rearrangement occurring at site $(ij)$. 
Besides, each node alternates between local plastic state ($n_{ij}=1$) and local elastic state ($n_{ij}=0$). The stochastic rule, as described in \cite{picard2005slow}, involves a rate of plastic activation $1/\tau_{pl}$ when the local stress exceeds a barrier $\mid\sigma_{ij}\mid>\sigma_{y}$ ($n_{ij}: 0 \rightarrow1 $)  and a rate $1/\tau_{el}$ for a plastic node turning elastic ( $n_{ij}: 1 \rightarrow 0$).
We consider in this work that a fluidizing noise induces additional plastic events ($n_{ij}: 0 \rightarrow1 $) with a ``vibration rate'' $k_{vib}=1/\tau_{vib}$, for any value of the local stress $\sigma_{ij}$.

In the following, the values of stress, strain rate and time are respectively given in units of $\sigma_{y}$, $\sigma_{y}/\mu\tau$ and $\tau$. 
We set $\tau_{pl}=1 $ and the restructuring time $\tau_{el}=10$ is chosen large compared to the other timescales in the system in order to induce local softening leading to non monotonic flow curves, as described in \cite{martens2012spontaneous}. 
We study the influence of an external noise by varying the value of the vibration rate $k_{vib}$ using both shear rate and stress controlled driving protocols, as they give access to different flow features in the case of non-monotonic flow curves.

\paragraph*{Flow transition at finite shear and vibration rates:}
 We measure the macroscopic flow curve of the system using a shear-rate-controlled protocol for different values of $k_{vib}$ controlling the magnitude of the fluidizing noise (Fig.\ref{fig1_Phenomenology}(a)). 
 The effect of the noise is
 (i) a vanishing yield stress at any value of $k_{vib}$ and (ii) a transition from a non-monotonic to a monotonic flow curve at a critical vibration rate $k_{vib}^{c}$ (thick black line in Fig.\ref{fig1_Phenomenology}(a)) associated with the transition from a shear banded flow \cite{martens2012spontaneous} to a homogeneous flow.

 Using a stress-controlled protocol with an imposed stress $\sigma_{ext}=\Sigma$, we examine the flow features starting from different initial states of the material: either flowing or arrested.
 In the negative slope region of the flow curve, this can lead to different values of the macroscopic shear rate even for large strains values \cite{SM}. 
 For a given value of the vibration rate $k_{vib}$, we estimate the minimum stress 
 value for which two flow solutions coexist within the time of our simulation (up to a strain $\gamma \simeq 800$). 
 
 We now introduce $S=\rm log_{10}(\dot\gamma)$, as in \cite{wortel2016criticality}. 
The two coexisting flow solutions $S_{slow}$ and $S_{fast}$ are depicted by the blue and red dots in Fig.\ref{fig1_Phenomenology}(b) for various values of $k_{vib}$ and delimit the regime of hysteretic flow (blue shadowed area in Fig.\ref{fig1_Phenomenology}(b)).
 The distance between the two branches $S_{fast}-S_{slow}$ then quantifies the ratio of the shear rates $\rm log_{10}(\frac{\dot\gamma_{fast}}{\dot\gamma_{slow}})$ in the two branches. 
 It decreases as vibration is increased, up to the point where it vanishes, consistent with the transition to a homogeneous flow in a shear-rate-controlled driving protocol (Fig.\ref{fig1_Phenomenology}(a)).
 This is reminiscent of equilibrium phase transitions, 
 where the distance between the two flow solutions  $\Delta S = S_{fast}-S_{flow}$ can be seen as the analogous of the density difference in the liquid-gas critical point.
 In the positive slope regions of the flow curve (red shadowed area in Fig.\ref{fig1_Phenomenology}(b)), there is a unique flow solution  for a given value of $\Sigma$ and $k_{vib}$. 
 We then measure the variance of $S$, $\Delta S^2$, in the regime where $k_{vib}>k_{vib}^c$ (Fig.\ref{fig1_Phenomenology}(c)) and find that, in the vicinity of the inflection point of the flow curve, decreasing $k_{vib}$ towards its critical value yields increasingly large fluctuations.

 Altogether, these observations suggest that this flow transition can be interpreted in the framework of non-equilibrium phase transitions, considering as an order parameter the distance between the two flow solutions  $\Delta S$, and the noise magnitude, denoted by the vibration rate $k_{vib}$, as control parameter (analogous of temperature), while the stress $\Sigma$ plays the role of the external field or pressure in equilibrium phase transitions.

 \paragraph*{Critical point analysis:}
We first investigate the scaling of the average value of $S$ with the imposed stress $\Sigma$ in the stable flow phase ($k_{vib}>k_{vib}^c$, monotonic flow curve). We find, as in \cite{wortel2016criticality}, that the data are well fitted to a Landau type expansion in the critical regime (Fig.\ref{fig2_flow_curve_scaling}(a)):
\begin{equation}
\Sigma=\Sigma_{i}+a\left(S-S_{i}\right)^{3}+b\left(S-S_{i}\right)
\label{fit_Landau}
\end{equation}
where $a$, $b$, $S_i$ and $\Sigma_i$ are fitting parameters. $a(k_{vib})$ is roughly constant (see SI) and $b$, which depends linearly on $k_{vib}$ close to the transition point (Fig.\ref{fig2_flow_curve_scaling}(b)), can be interpreted as an inverse susceptibility, $b=1/ \chi$.

To get the critical vibration rate $k_{vib}^c$, we estimate the value of $k_{vib}$ corresponding to $b(k_{vib})=0$ for each system size $L$ (Fig.\ref{fig2_flow_curve_scaling}(b)), and perform a linear extrapolation to get the critical value in the limit of an infinite system size (Fig.\ref{fig2_flow_curve_scaling}(c)), leading to $k_{vib}^c = (1.35 \pm 0.01) 10^{-3}$. 

Moreover, the agreement with a linear fit in Fig.\ref{fig2_flow_curve_scaling}(c) suggests that the value of the exponent $\nu$ related to the correlation length $\xi$ should be close to 1, as one would expect a scaling for the shift of the critical rate of the form:
$|k_{vib}^c(L)-k_{vib}^c(\infty)| \sim L^{-1/\nu}$ \cite{binder2010monte}.

To verify this scaling, we measure the correlation length of the system $\xi$ (for $L=1024$) from the Fourier transform of the autocorrelation of the plastic deformation rate field, $\tilde C(q)$ for various values of $k_{vib}$ \cite{SM}. 
In the vicinity of a critical point, we expect a power law scaling of the form: $\tilde C(q) \sim q^{\alpha}$ \cite{LeBellacBook}. We estimate the value of $\alpha$ from data where $k_{vib} \simeq k_{vib}^c $ (see \cite{SM}, $\alpha \simeq 0.75$). 
We show, in Fig.\ref{fig2_flow_curve_scaling}(d), $\tilde C(q) q^\alpha$ as a function of $q \xi$, with $\xi = \bigl((k_{vib}-k_{vib}^c)/{k_{vib}^c}\bigr)^{\nu}$. The best collapses are found for $\nu \in [0.9;1]$, consistent with the linear scaling of Fig.\ref{fig2_flow_curve_scaling}(c).


\begin{figure}[t]
\centering
\includegraphics[width=1.0\columnwidth, clip]{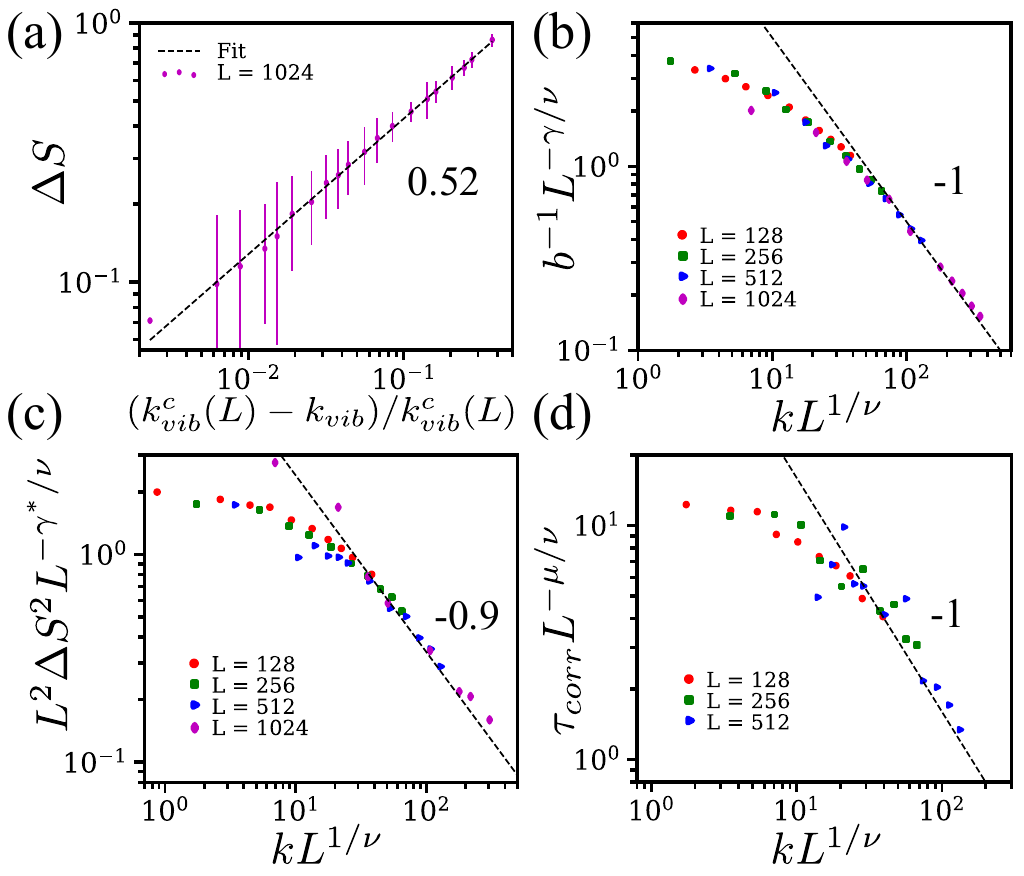}
\caption{ \textbf{(a) Order parameter} $\Delta S = S_{fast}-S_{slow}$ as a function of $(k_{vib}-k_{vib}^c(L))/k_{vib}^c(L)$ for $L = 1024$ and power law fit (see text).  
\textbf{(b) Susceptibility.} $\chi = 1/b$ from the fit of Fig.\ref{fig2_flow_curve_scaling}(a-b) as a function of $kL^{1/\nu}$ with $k = |k_{vib}-k_{vib}^c|/k_{vib}^c$.
\textbf{(c) Magnitude of order parameter fluctuations} $L^2 \Delta S ^2$ as a function of $kL^{1/\nu}$.
\textbf{(c) Temporal correlations of fluctuations:} characteristic time $\tau_{corr}$ as a function of $kL^{1/\nu}$}
\label{fig3_fluctuations}
\end{figure}

We now examine the scaling of the order parameter $\Delta S$ with the distance to the critical point, by performing a power law fit of the form:
$\Delta S(L)= A (( k_{vib}^c(L)-k_{vib})/k_{vib}^c(L) )^{\beta}$.
In Fig.\ref{fig3_fluctuations}(a), we find a good agreement with a power law with the following parameters for $L=1024$: $k_{vib}^c(L) = (1.324 \pm 0.005) 10^{-3}$, $A=1.43 \pm 0.03$ and $\beta = 0.52 \pm 0.02$. 
Note that the value of $k_{vib}^c(L)$ is consistent with the value obtained from the divergence of the susceptibility ($1/\chi = b(k_{vib})=0$) in Fig.\ref{fig2_flow_curve_scaling}(b) (where we had, for $L=1024$,  $k_{vib}^c(L) = (1.33 \pm 0.02) 10^{-3}$). 

Using the above results for the correlation length critical exponent $\nu \simeq 1$, we can now test finite size scaling of the different quantities measured in our simulations: the susceptibility $\chi = 1/b $ (Fig.\ref{fig3_fluctuations}(b)), as well as the variance $\Delta S ^2$ (Fig.\ref{fig3_fluctuations}(c)) and correlation time $\tau_{corr}$ (Fig.\ref{fig3_fluctuations}(d)) of the order parameter fluctuations.

As shown in Fig.\ref{fig3_fluctuations}(b), we can indeed collapse the susceptibility data obtained for the different system sizes (from Fig.\ref{fig2_flow_curve_scaling}(b)) using $\nu = 1$, leading to a power law increase as a function of the scaled distance to the critical point, $k = |k_{vib}-k_{vib}^c|/k_{vib}^c$, with an exponent $\gamma \simeq 1$.
As seen in Fig.\ref{fig1_Phenomenology}(c), the fluctuations of the order parameter exhibit a maximum at the inflection point of the flow curve (as well as the correlation time \cite{SM}), and we represent the maximum of $\Delta S ^2$ and $\tau_{corr}$ in Fig.\ref{fig3_fluctuations}(c) and (d).
The increase of the variance $\Delta S ^2$ of fluctuations when approaching the critical point can be well described with a power-law and the best collapse is found for $\gamma^* \simeq 0.9$.
The data for the correlation time, determined from an exponential fit of the temporal autocorrelation function of $S$ \cite{SM}, appear quite noisy due to finite-time limitations of our simulations in the critical region, but finite size data collapse can still be performed, with a power-law scaling with an exponent $\mu \in [0.8;1.2]$. This corresponds to a dynamic scaling exponent $z=\mu/\nu \approx 1$, far from the equilibrium mean-field value $z=2$ obtained for non-conserved scalar order parameters \cite{HohenbergRMP1977}.

\begin{table}[t]
\small
   \centering
\begin{tabular}{|c|c|c|c|c|c|}
\hline
Scaling relation & Exp. & This work & Wortel et al. \cite{wortel2016criticality} & MF\\
\hline
$\Sigma-\Sigma_c \sim (S - S_c)^{\delta}$  & $\delta$ &3 & 3 & 3  \\
\hline
$\xi \sim k^{-\nu}$ & $\nu$ & $1.0 \pm 0.1$ & - & 0.5   \\
\hline
$\Delta S \sim k^{\beta}$ & $\beta$ & $0.52 \pm 0.02$& 0.5 & 0.5  \\
\hline
$\chi \sim k^{-\gamma}$  & $\gamma$ & $1.0 \pm 0.05$ & 1 & 1  \\
\hline
$\Delta S^2 \sim k^{-\gamma^*}$ & $\gamma^*$ & $0.9 \pm 0.05$ & 1 & 1  \\
\hline
$\tau_{corr} \sim k^{-\mu}$ &$\mu$ & $1.0 \pm 0.2$ & $[0.5;1]$& 0.5   \\
\hline
\end{tabular}
   \caption{Critical exponents measured in our model, compared with the values obtained in Ref.~\cite{wortel2016criticality} and with standard equilibrium mean-field values (MF).}
   \label{table_exponents}
\end{table}

 \paragraph*{Discussion:}
The above analysis shows evidences for the existence of a critical point at finite strain rate in a mesoscopic model for the flow of soft glassy materials with an external noise, characterized by the scaling exponents summarized in Table \ref{table_exponents}.
We located the critical point $k_{vib}^c = (1.35 \pm 0.01) 10^{-3}$ and checked that this value was consistent between different independent measurements (divergence of susceptibility, finite size data collapse and order parameter fitting). 
The critical exponent of the susceptibility, obtained from an average quantity, $S$, ($\gamma \simeq 1$) is found to be close to that of the fluctuations, $\Delta S^2$, ($\gamma^* \simeq 0.9$) but not identical. 
This difference may be explained by the fact that fluctuations are slightly underestimated in our analysis.

The quality of the flow curve fits with a Landau-type expansion, as well as the values of some of the exponents ($\beta$, $\gamma$, $\gamma^*$), indicate that the average quantities scaling is close to a standard mean field scaling for equilibrium phase transitions, as observed by Wortel et al. \cite{wortel2016criticality}.
Interestingly, the scaling of the correlations, however, departs from standard exponent values \cite{LeBellacBook,HohenbergRMP1977} and are again consistent, within error bars, with the values obtained in \cite{wortel2016criticality}. 
In conclusion, the critical point studied here in a frictionless model exhibits similar critical properties as the one studied experimentally in a frictional system \cite{wortel2016criticality}, suggesting that a generic critical behavior arises in systems combining a non-monotonic flow curve with a fluidization process, irrespective of the detailed physical mechanisms at play.
In addition, the critical exponents characterizing the divergence of correlation length and times take non-standard values, that cannot be easily inferred from an equilibrium analogy, even taking into account long range interactions \cite{SM}.
It would be of interest to confirm this generic scenario in different types of experiments, where the physical origin of the non-monotonic flow curve and of the fluidization mechanism could be varied.

\begin{acknowledgments}
KM and MLG acknowledge funding from the Centre Franco-Indien pour la Promotion de la Recherche Avanc\'ee  (CEFIPRA) Grant No.\ 5604-1 (AMORPHOUS-MULTISCALE). KM acknowledges financial support of the French Agence Nationale de la Recherche (ANR), under grant ANR-14-CE32-0005 (FAPRES).
Further the authors would like to thank Olivier Dauchot, Vishwas Vasisht and Vivien Lecomte for valuable discussions about this work.
\end{acknowledgments}

\bibliography{StageM2.bib}

\pagebreak

\begin{widetext}

\begin{center}
\textbf{Supplemental Material for:\\
``Criticality at finite strain rate in fluidized soft glassy materials''}
\end{center}
\end{widetext}

\setcounter{equation}{0}
\setcounter{figure}{0}
\setcounter{table}{0}

\subsection*{Hysteretic flow regime:}

Using a stress-controlled driving protocol we examine the hysteretic flow regime (corresponding to shear banded flow when using a shear rate controlled protocol), where the flowing state reached even for large strain amplitudes can depend on the initial conditions. 
Fig.1 depicts two examples of $S=\rm log(\dot\gamma)$ as a function of time starting from either flowing or arrested states, in the hysteretic region (panel (a)) and near the critical point (panel (b)), for strain values up to 800.
The order parameter is defined as the difference between the two flowing branches $S_{fast}-S_{slow}$ at large strain, with a non-zero value in the hysteretic regime (panel (a)), and going towards 0 approaching the critical point (panel (b)).
The value of stress $\Sigma$ selected to measure the order parameter is chosen such that it is the lowest value of stress for which the fast flowing branch remains stable within the time of our simulation (for strain values $\gamma \in [700;1000]$).
This method gives a robust measurement of the scaling of the order parameter in the phase separation regime, although it can seem somehow arbitrary because it doesn't give a direct access to the exact binodal or spinodal lines of the system.

%

\begin{figure}[h]
\centering
\includegraphics[width=1.0\columnwidth, clip]{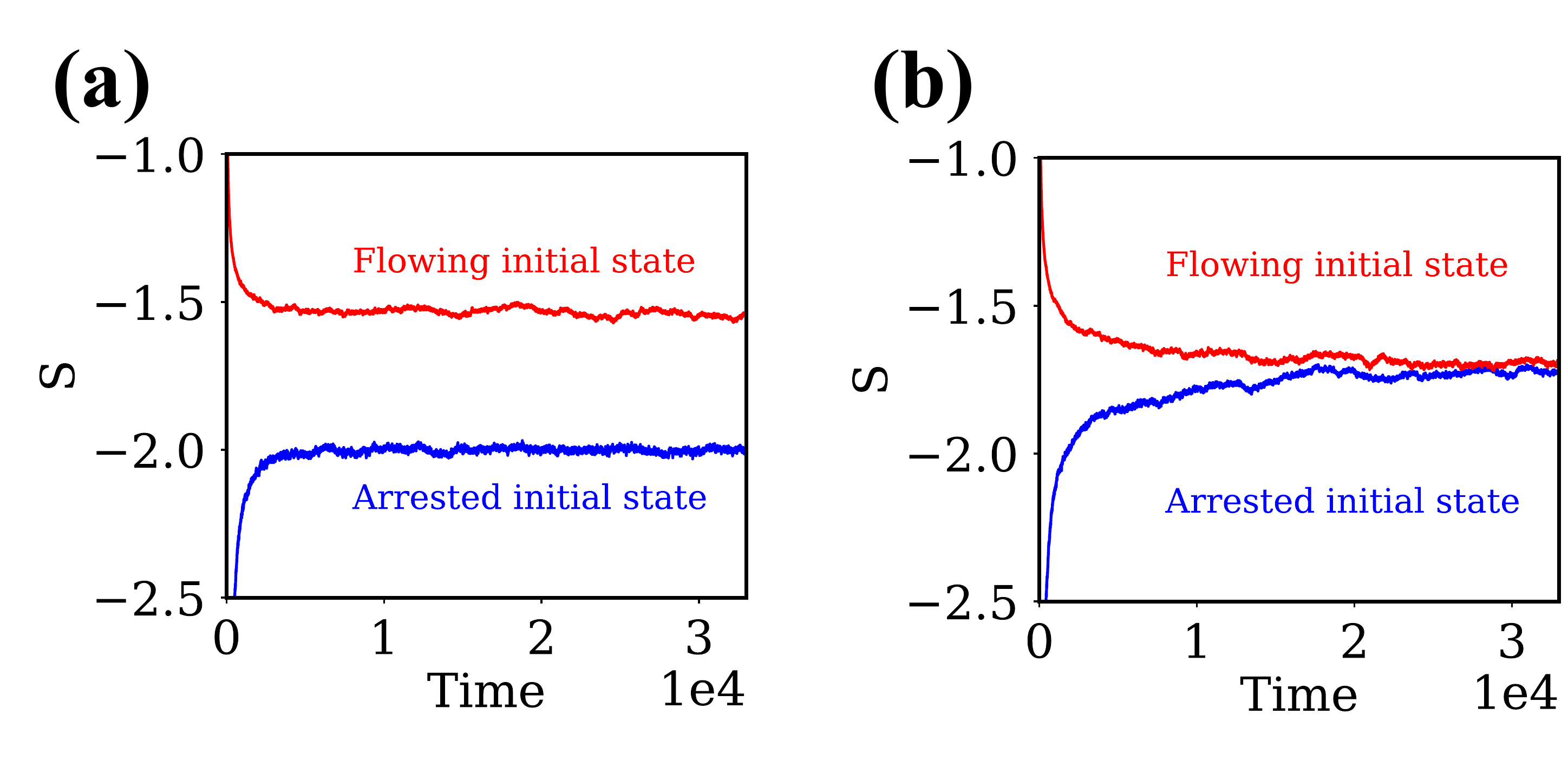}
\caption{\textbf{Hysteretic flow regime} $S=\rm log(\dot\gamma)$ as a function of time starting from either a flowing (red) or an arrested state (blue) (system size $L=1024$): \textbf{(a) Phase separated regime} with $k_{vib} = 1.177.10^{-3}$ and $\Sigma = 0.47715$ \textbf{(b) Near the critical point} with $k_{vib} = 1.324.10^{-3}$ and $\Sigma = 0.47595$ }
\label{fig1_supp}
\end{figure}
%

\subsection*{Flow curve fitting}
The stress $\Sigma$ as a function of the average value of $S=\rm log(\dot\gamma)$ is well fitted with the following equation (see main text, Fig.2(a)): 
\begin{equation}
\Sigma=\Sigma_{i}+a\left(S-S_{i}\right)^{3}+b\left(S-S_{i}\right)
\label{fit_Landau}
\end{equation}
where $a$, $b$, $S_i$ and $\Sigma_i$ are free fitting parameters. The value of $b$ is displayed in the main text in Fig.2(b), and the values of $a$, $S_i$ and $\Sigma_i$ are depicted in Fig.2 as a function of $k_{vib}$ in the regime $k_{vib}>k_{vib}^c$ (stable flow regime).
$a$ varies only slightly with $k_{vib}$, although there seem to be some finite size effects which could be further discussed.
$S_i$ and $\Sigma_i$ vary monotonically as $k_{vib}$ is increased.


\begin{figure}[h]
\centering
\includegraphics[width=0.9\columnwidth, clip]{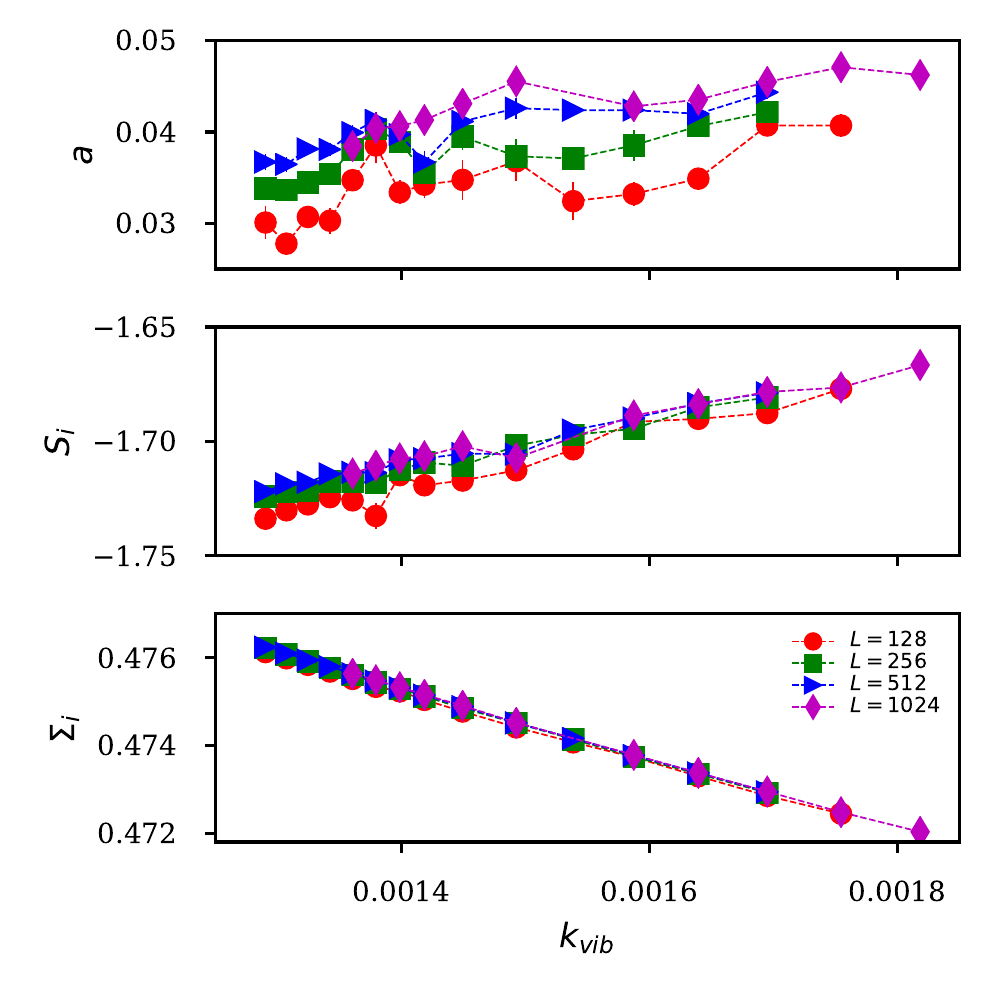}
\caption{\textbf{Fitting parameters of the flow curve as a function of $k_{vib}$}. \textbf{Top}: $a$. \textbf{Middle and bottom:} Coordinates of the inflexion point of the flow curve, $S_i$ (middle) and $\Sigma_i$ (bottom)}
\label{fig2_supp}
\end{figure}

\subsection*{Spatial correlations:}

In order to determine how spatial correlations evolve in the system as $k_{vib}$ is varied, we compute the squared modulus of the fourier transform of the instantaneous local shear rate configuration, averaged over at least 10 000 configurations, $\tilde C(q)$. 
$\tilde C(q)$ is depicted in Fig.3 for various values of $k_{vib}$. 
Note that the data look noisier as approaching $k_{vib}^c$, due to growing time correlations in the system near the critical point, i.e. due to finite-time limitations of our simulations.
In the vicinity of the critical point, we expect a scaling of the form $\tilde C(q) \sim q^\alpha$. From the thick solid line of Fig.3, we get $\alpha \simeq 0.75$.

\begin{figure}[h!]
\centering
\includegraphics[width=0.7\columnwidth, clip]{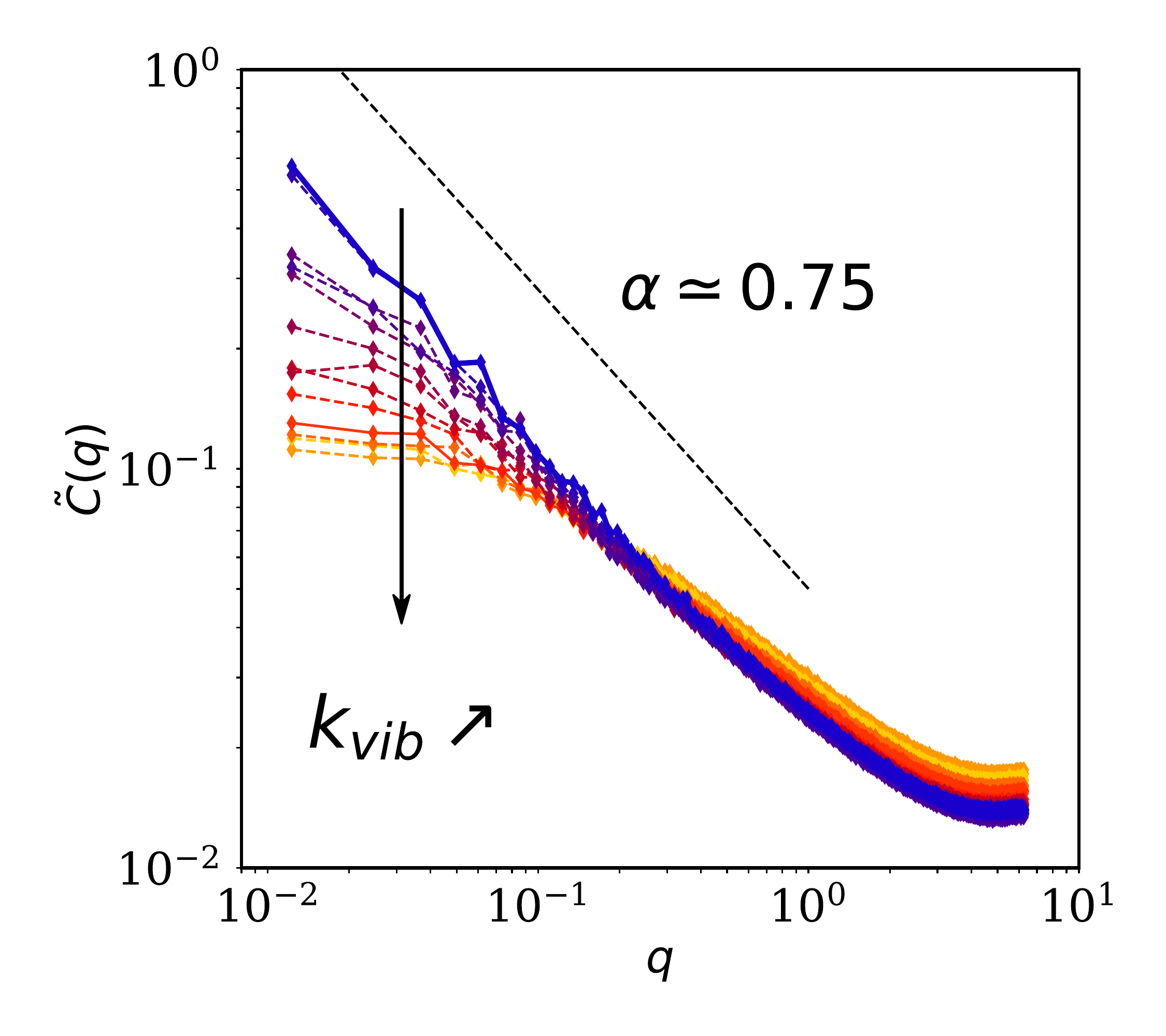}
\caption{\textbf{Spatial correlations}  $\tilde C(q)$ as a function of $q$ for a system size $L=1024$ and various values of $k_{vib}$ ranging from $k_{vib}^c=1.35.10^{-3}$ (solid blue line) to $k_{vib}=1.81.10^{-3}$ (yellow dashed line)}
\label{fig3_supp}
\end{figure}

\begin{figure}[h!]
\centering
\includegraphics[width=1.0\columnwidth, clip]{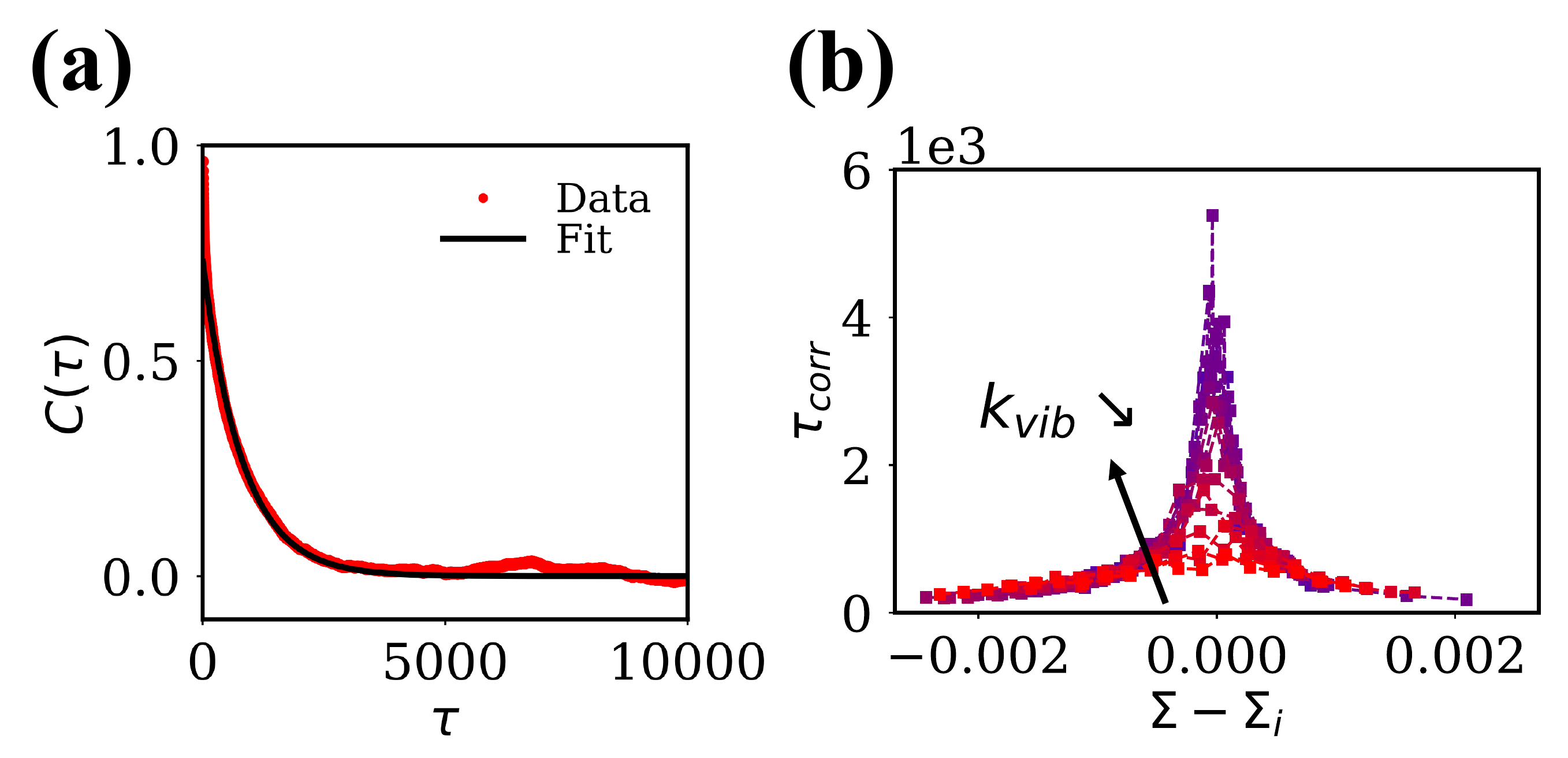}
\caption{\textbf{Temporal correlations}  \textbf{(a)$C(\tau)$ as a function of $\tau$.} Example shown for $k_{vib}=1.42.10^{-3}$, $\Sigma=0.4746$ and $L=256$. Reds dots represent autocorrelation data and black solid line represent the exponential fit used to extract the autocorrelation timescale $\tau_{corr}$.  \textbf{(b) Autocorrelation timescale $\tau_{corr}$} as a function of $\Sigma-\Sigma_i$ for various vibration rates $k_{vib}$.}
\label{fig4_supp}
\end{figure}


\subsection*{Details on data analysis:}
The average value of $S$, the variance $\Delta S^2$ of the order parameter fluctuations and their correlation time $\tau_{corr}$ are extracted from time-series of the flow rate (of average duration $T=2.10^6$ for $L=128, 256$, $T=6.10^5$ for $L=512$ and $T=10^5$ for $L=1024$, corresponding to strains ranging from $\gamma=2000$ to $4.10^4$.

The correlation time $\tau_{corr}$ is extracted by fitting the autocorrelation function 
$C(\tau)=\langle \Delta S(t+\tau)\Delta S(t)\rangle$ 
to an exponential, as depicted in Fig.4(a), 
and exhibit a sharp peak corresponding to the inflexion point of the flow curve, increasingly large as approaching the critical point, as shown in Fig.4(b).

\hspace*{0mm}

\subsection*{Analogy with equilibrium critical phenomena with long-range interactions}

Given that the exponents found for the correlation length and time are different from the equilibrium mean-field exponents for systems with short-range interactions, it is natural to wonder if including long-range interactions in an equilibrium analogue of our model may lead to the exponents $\nu=1$ and $\mu=1$.
For the sake of simplicity, we briefly discuss this issue here in the language of spin models, where a magnetization field $m(\mathbf{r})$ is introduced. The above values of the exponents are suggestive of an effective Hamiltonian of the form (in the Gaussian approximation)
\begin{equation} \label{eq:SI:Hamilt:eff}
    H \propto \int d\mathbf{q} \int d\mathbf{q}' (\varepsilon + |\mathbf{q}|) \hat{m}(\mathbf{q}) \hat{m}(-\mathbf{q})
\end{equation}
where $\varepsilon$ is the dimensionless deviation from the critical point, and $\hat{m}(\mathbf{q})$ is the spatial Fourier transform of the field $m(\mathbf{r})$. Such a Gaussian form leads to a divergence of the correlation length $\xi \sim \varepsilon^{-1}$, and thus to $\nu=1$. However, this form corresponds to interactions decaying as $1/r^{d+1}$ (where $d$ is the space dimension, $d=2$ in our model), and not as $1/r^d$ as the Eshelby propagator.

Note that in terms of dynamics, a simple Langevin relaxational dynamics with the effective Hamiltonian (\ref{eq:SI:Hamilt:eff}) would lead, at the critical point, to
\begin{equation}
    \partial_t \hat{m} = - |\mathbf{q}| \hat{m}(\mathbf{q}) + \xi(\mathbf{q},t)
\end{equation}
with $\xi(\mathbf{q},t)$ a white noise.
At a heuristic level, the scaling `time $\sim$ length' suggests a dynamical exponent $z=1$, corresponding to $\mu=z\nu=1$.

\end{document}